\documentclass[5p]{elsarticle}
\usepackage{amsmath}
\usepackage{amssymb}
\usepackage{graphicx}

\newcommand{\be}{\begin{equation}}
\newcommand{\ee}{   \end{equation}}
\newcommand{\ve}{\varepsilon}

\begin{document}

\title{Photonuclear Reactions induced by Intense Short Laser Pulses}

\author{B. Dietz\fnref{ikp}} 

\author{H. A. Weidenm{\"u}ller\corref{cor}\fnref{mpi}}
\ead{Hans.Weidenmueller@mpi-hd.mpg.de}

\address[ikp]{Institut f{\"u}r Kernphysik, Technische Universit{\"a}t
Darmstadt, D-64289 Darmstadt, Germany}

\address[mpi]{Max-Planck-Institut f{\"u}r Kernphysik, D-69029 Heidelberg,
Germany}

\cortext[cor]{Corresponding author}

\date{\today}

\begin{abstract}
A measurement of the decay in time of nuclei excited by an intense
short laser pulse of energy $E_0$ yields the Fourier transform of the
autocorrelation function of the associated scattering matrix. We
determine the optimal length (in time) of the pulse and evaluate the
time--decay function using random--matrix theory. That function is
shown to contain information not otherwise available. We approximate
that function in a manner that is useful for the analysis of data. For
$E_0$ below the threshold energy $E_n$ of the first neutron channel,
the time--decay function is exponential in time $t$ while it is the
product of an exponential and a power in $t$ for $E_0 > E_n$. The
comparison of the measured decay functions in both energy domains
yields an unambiguous and novel test of random--matrix theory in
nuclei.
\end{abstract}

\maketitle

\section{Purpose}

ELI, the ``Extreme Light Infrastructure'', an ambitious European
project to generate laser beams of extremely high intensity, is close
to construction. Parallel to that development, experiments that will
use those beams are being planned, and theorists are called upon to
develop the concepts and tools needed for their analysis. At the
workshop on ELI held in Palaiseau (France) April 27/28, 2009,
photonuclear reactions induced by an intense laser pulse of high
energy (several MeV) received much attention, see Ref.~\cite{Sac09}.
The required high--energetic directed pulsed gamma rays are supposed
to be produced by Compton backscattering of a short laser pulse with
much lower energy on a sheet of electrons ejected from a thin foil hit
by the ELI pulse.

In this paper, we present a theoretical study of nuclear reactions
induced by short laser pulses of several MeV energy. In particular, we
address the following questions: (i) Which observable is measured in a
photonuclear reaction induced by a very short laser pulse? (ii) Which
novel information is provided by data of that type? (iii) What is the
optimal length in time of a short laser pulse for such a reaction?

We focus attention on the main mode of nuclear excitation by gamma
quanta of several MeV energy, the electric dipole mode. For a target
nucleus in its ground state with spin $J$ and parity $\pi$, dipole
absorption leads to excited states with spin $J \pm 1$ and opposite
parity $- \pi$. The analysis of experimental data will be simplest
when the spins of the excited states are uniquely defined. Therefore,
we consider an even--even target nucleus with a ground state of spin
zero and positive parity. Then the states excited by dipole absorption
have spin $1$ and negative parity. Depending on excitation energy,
these may lie below or above the first particle threshold. That is
typically the threshold for neutron emission. In medium--weight and
heavy nuclei, that threshold has an excitation energy of 5 to 8
MeV. Nuclear states right above neutron threshold have been studied by
time--of--flight spectroscopy in slow neutron scattering, mostly on
even--even target nuclei. Such states appear as isolated $s$--wave
resonances with spin $1/2$ and positive parity, with a typical spacing
of $10$ eV and a typical width of $1$ eV. The statistical analysis of
such resonances shows that the spacings and widths follow the
predictions of random--matrix theory (RMT), see the
review~\cite{Wei09}. In recent years, states in even--even nuclei with
spin $1$ and negative parity close to the first particle threshold
forming the ``pygmy dipole resonance'' have been intensely studied
experimentally, mainly with the help of the resonance fluorescence
technique, see Refs.~\cite{Rye02,End04} and references therein. In
Ref.~\cite{End04} deviations from RMT predictions were found that have
so far not been fully explained theoretically. Our work is based on
the assumption that the states excited by an intense short laser pulse
are governed by RMT. We explore the consequences of that assumption
and propose an experimental test for it.

For reasons given below, we consider a laser pulse of $10^{-19}$ to
$10^{- 20}$ s duration and a mean energy of several MeV. The pulse
coherently excites a band of $1^-$ states. The band width is several
$10$ keV, the number of $1^-$ states involved is typically $10^3$ to
$10^4$. We refer to these states as to compound--nucleus (CN)
resonances. These will subsequently decay. CN resonances below neutron
threshold decay by gamma emission, those above neutron threshold
preferentially by neutron emission. Gamma emission is possible but
less likely. In both cases the detection of the emitted particle with
highest energy unambiguously identifies decay into the ground state of
the residual nucleus. We do not focus attention solely on that decay
mode since in the case of gamma decay, the intensity may be too low,
and a summation of intensities corresponding to several or many gamma
decay modes leaving the target nucleus in its ground or one of its
excited states, may be called for. In any case we deal with a
two--channel situation. The incident channel $\gamma_0$ is defined by
the target nucleus in its ground state plus a dipole gamma quantum,
the exit channel $b$ by the emitted particle ($b = n$ for neutron
emission leaving the residual nucleus in its ground state, $b =
\gamma_i$ with $i = 0, 1, \ldots, \Lambda$ for gamma emission leaving
the residual nucleus in its ground or any of $\Lambda$ excited
states).

Formation and decay of the CN resonances being coherent processes, it
is not possible to identify any particular CN resonance as the source
of the emitted particle. Interest in the data rather focuses on the
time dependence of the decay. We answer the questions raised in the
second paragraph by identifying the relevant observable and defining
the optimal length in time of the laser pulse. We also work out the
expected form of the time--decay function. We do so by using results
of the analytical approach to CN reactions developed in
Ref.~\cite{Ver85}.

Coherent resonance formation was previously addressed, see
Ref.~\cite{Rem90} and references therein, the decay in time of CN
resonances in Refs.~\cite{Har90,Dit92,Har92}. The deviations from the
exponential decay in time predicted in Refs.~\cite{Har90,Dit92,Har92}
have been observed in microwave billiards by Fourier--transforming the
measured elements of the scattering matrix, see
Refs.~\cite{Alt95,Mit10}, and the comprehensive summary
paper~\cite{Die10a}. Here we propose a direct measurement of the
time--decay function in nuclei. We go beyond
Refs.~\cite{Har90,Dit92,Har92} in addressing specifically the case of
photonuclear reactions induced by an intense laser pulse. In
Ref.~\cite{Fyo98} the autocorrelation function of the total
photodissociation cross section for a chaotic atom or molecule was
studied. With the help of the optical theorem, that function is
related to a two--point function similar in form to the expressions
studied below. The approach was extended and generalized in
Refs.~\cite{Gor04,Gor05}.

\section{Observable}

To describe the effect of a short light pulse, we consider the
scattering wave function $\Psi^+_{\gamma_0}(E)$. That function
describes all reactions caused by a gamma quantum of energy $E$ with
wave number $k(E)$ incident on the target nucleus. Asymptotically,
$\Psi^+_{\gamma_0}(E)$ has an incident wave with unit flux in channel
$\gamma_0$ and outgoing waves in all channels. The amplitude of the
outgoing flux in channel $b$ is given by the element $S_{b
\gamma_0}(E)$ of the scattering matrix, and the cross section
$\sigma_{b \gamma_0}(E)$ feeding final channel $b$ is given by
\be
\sigma_{b \gamma_0}(E) = \frac{\pi}{k^2} |S_{b \gamma_0}(E) -
\delta_{b \gamma_0}|^2\, .
\label{17}
\ee
A short light pulse is described as a superposition 
\be
\int {\rm d} E \ g(E) \exp \{ - i E t / \hbar \} \Psi^+_{\gamma_0}(E) 
\nonumber
\ee
of scattering
wave functions with different energies where $t$ denotes the time. The
envelope function $g(E)$ is a smooth function of energy $E$ centered
at energy $E_0$ (the mean energy of the laser pulse), has band width
$\Delta E$, and $|g(E)|^2$ is normalized to unity. The amplitude of
the outgoing flux in channel $b$ is $\int {\rm d} E \ g(E) \exp \{ - i
E t / \hbar \} S_{b \gamma_0}(E)$. The total scattered flux $I_{b
\gamma_0}(t)$ in channel $b$ versus time is given by
\be
\frac{I_{b\gamma_0}(t)}{\pi}=\bigg|\int{\rm d}E\ \frac{g(E)}{k(E)}
\exp\{-iEt/\hbar\}[S_{b\gamma_0}(E)-\delta_{b\gamma_0}]
\bigg|^2\, .
\label{1}
\ee
Indeed, integrating the total scattered flux over time and using
Eq.~(\ref{1}) we obtain the energy--averaged cross section, $\int {\rm
d} t / h \ I_{b \gamma_0}(t) = \int {\rm d} E \ |g(E)|^2 \sigma_{b
\gamma_0}(E)$. In other words, $I_{b \gamma_0}(t)$ gives the
decomposition in time of the cross section induced by a laser pulse of
band width $\Delta E$ and time length $\Delta t = \hbar / \Delta E$. A
short laser pulse is defined to have a band width $\Delta E$ that is
large compared to the average spacing $d$ of the CN resonances. We
focus attention on times $t$ that are large compared to $\Delta t$ and
on the associated long--time behavior $I^{\rm long}_{b \gamma_0}(t)$
of $I_{b \gamma_0}(t)$. We do so because it may be difficult to
separate experimentally short--time contributions from the original
laser signal, and we accordingly expect that the first round of
experiments will focus on the long--time aspects of the reaction. The
terms in Eq.~(\ref{1}) that involve $\delta_{b \gamma_0}$ are
proportional to the Fourier transform of $g(E)$, have time length
$\Delta t$, and are, therefore, neglegted. Thus,
\begin{align}
\label{1a}
&I^{\rm long}_{b \gamma_0}(t) \\ 
&=\frac{\pi}{k^2_0} \bigg| \int
{\rm d} E \ g(E) \exp \{ - i E t / \hbar \} S_{b \gamma_0}(E) \bigg|^2
\nonumber \\
&=\frac{\pi}{k^2_0} \int {\rm d} E_1 \int {\rm d} E_2 \ g(E_1)
g^*(E_2) \exp \{ i (E_2 - E_1) t / \hbar \} \nonumber \\
&\times S_{b \gamma_0}(E_1) S^*_{b \gamma_0}(E_2) \, .
\nonumber
\end{align}
Here and in what follows we use that $k(E)$ changes little over the
energy interval $\Delta E$ and replace $k(E)$ by $k_0 = k(E_0)$. (We
note that $I^{\rm long}$ still contains a short--time component, see
Eq.~(\ref{10}) below. That component will also be suppressed
eventually). We relate $I^{\rm long}_{b \gamma_0}(t)$ to the
$S$--matrix autocorrelation function $C_{b \gamma_0}(\ve)$ defined by
\be
C_{b \gamma_0}(\ve) = \int {\rm d} E \ |f(E)|^2 S_{b \gamma_0}(E -
\ve / 2) S^*_{b \gamma_0}(E + \ve / 2) \, .
\label{2}
\ee
The average is performed with a smooth function $|f(E)|^2$ normalized
to unity the range of which extends to infinity. An example is a
normalized Lorentzian with a width that is very large compared to $d$.
We deal with isolated CN resonances for which the average total
resonance width $\Gamma$ obeys $\Gamma < d$. Then the function $C_{b
\gamma_0}(\ve)$ decreases rapidly towards zero for $|\ve| > d$. In
Eq.~(\ref{1a}) we write $E_1 = E - \ve / 2$, $E_2 = E + \ve / 2$. We
use that $g(E)$ is smooth over the correlation width of $C_{b
\gamma_0}(\ve)$. Then $g(E \pm \ve / 2) \approx g(E)$, and
Eq.~(\ref{1a}) becomes
\begin{align}
\label{3}
&I^{\rm long}_{b \gamma_0}(t) \\ 
&\approx\frac{\pi}{k^2_0} \int {\rm d}
\ve\,\exp \{ i\ve t/\hbar \} \, 
\nonumber\\
&\times\int{\rm d} E\, |g(E)|^2 S_{b
\gamma_0}(E - \ve / 2) S^*_{b \gamma_0}(E + \ve / 2) \nonumber \\
&\approx\frac{\pi}{k^2_0}\int{\rm d}\ve\exp\{ i\ve t/\hbar
\}C_{b \gamma_0}(\ve)\, .
\nonumber
\end{align}
The approximation leading to Eq.~(\ref{3}) requires $\Delta E \gg
d$. Otherwise, finite--range--of--data errors have to be taken into
account. If that condition is met, the long--time part of the signal
$I_{b \gamma_0}(t)$ measures the Fourier transform of the $S$--matrix
autocorrelation function $C_{b \gamma_0}(\ve)$. The inequality $\Delta
E \gg d$ yields the first constraint on the length in time of the
laser pulse: $\Delta t$ must be small compared to the Heisenberg time
$\hbar / d$.

A second constraint arises because dipole absorption in nuclei is
governed by the giant dipole resonance. Depending on mass number the
resonance occurs at excitation energies between $10$ and $15$ MeV and
has a width $\Gamma^{\rm dip}$ of several MeV. The resonance causes a
secular variation of the average dipole strength of the CN
resonances. For a clean theoretical analysis, it is desirable to
separate that secular variation from the average over resonances taken
in the first of Eqs.~(\ref{3}). That condition is met if $\Delta E \ll
\Gamma^{\rm dip}$.

Combining the two constraints we find $d \ll \Delta E \ll \Gamma^{\rm
dip}$. With $d \approx 10$ eV and $\Gamma^{\rm dip} \approx$ several
MeV, the constraints are met for $\Delta E$ in the range $10$ keV to
$100$ keV, depending on excitation energy and mass number. That
corresponds to a time length $\Delta t \approx 0.5 \times 10^{-19}$ s
to $0.5 \times 10^{-20}$ s. Under these conditions, irradiation of a
target with an intense laser pulse yields information on the decay in
time of the CN resonances. As shown in Refs.~\cite{Dit92,Har92}, that
decay is not expected to be exponential in general. It would certainly
be exciting to measure directly the time--dependence of the decay of
the CN, i.e., of the signal $I_{b \gamma_0}(t)$, see
Section~\ref{auto}, even though measurements in microwave billiards
have already provided clear evidence for non--exponential decay in
these systems~\cite{Alt95,Mit10,Die10a}. Such measurements are
possible because the average width for neutron decay is less than $1$
eV and that for gamma decay even smaller. The overall decay time of
the CN resonances is, thus, much longer than the duration of the
incident laser pulse, and the two signals are clearly separated.

We note that a band width $\Delta E \approx 10 \ldots 100$ keV also
guarantees that in exciting CN resonances right above neutron
threshold, one avoids excitation energies where neutron emission
leaving the residual nucleus in an excited state becomes possible.
(The excitation energy of the first excited state in the residual
nucleus is typically $100$ to $200$ keV).  This is desirable as
otherwise in the expression for $C_{b \gamma_0}(\ve)$ the additional
neutron channel must be taken into account, even if only decay into
the ground state is experimentally measured.

The decay in time of CN resonances excited by a short and intense
laser pulse is described by the Fourier transform of the $S$--matrix
autocorrelation function $C_{b \gamma_0}(\ve)$. Is that information
novel, or are other data available or within experimental reach that
would yield the same information? An obvious possibility would be, for
instance, a measurement of the $(\gamma_0, n)$ reaction cross section
$\sigma_{n, \gamma_0}(E)$ from which the cross--section
autocorrelation function 
\be
C^\sigma_{n, \gamma_0}(\ve) = \int {\rm d} E
\ |f(E)|^2 \sigma_{n, \gamma_0}(E + \ve / 2) \sigma_{n, \gamma_0}(E -
\ve / 2)\, , 
\nonumber
\ee
with $f(E)$ as defined in Eq.~(\ref{2}) can be obtained.
For isolated resonances, $C^\sigma_{n, \gamma_0}(\ve)$ is neither
theoretically accessible (in contrast to $C_{b \gamma_0}(\ve)$, see
Section~\ref{auto}), nor is it simply related to $C_{b
\gamma_0}(\ve)$, see Ref.~\cite{Die10}. Generally speaking, the
coherent decay in time of the CN resonances is described by an
amplitude correlation function. An intensity autocorrelation function
such as $C^\sigma_{n, \gamma_0}(\ve)$ does not provide equivalent
information (except in the Ericson regime where $\Gamma \gg d$). Thus,
the information available from $C_{b \gamma_0}(\ve)$ is unique.

\section{$S$--Matrix Autocorrelation Function}
\label{auto}

The autocorrelation function for CN scattering has been calculated
analytically for CN resonances that obey RMT statistics~\cite{Ver85}.
In Ref.~\cite{Die08} and references therein it was shown that this
function correctly describes chaotic scattering. Without going into
details, we summarize some salient features of that approach, see
Refs.~\cite{Mit10,Die10a}, and cite the result. The $S$-matrix with
elements $S_{b a}(E)$ is decomposed into an average part $\langle S_{b
a} \rangle$ and a fluctuating part $S^{\rm fl}_{b a}(E)$,
\be
S_{b a}(E) = \langle S_{b a} \rangle + S^{\rm fl}_{b a}(E) \, .
\label{4}
\ee
The average is taken over the ensemble of random matrices. In what
follows we use the equality of ensemble average and energy average
which holds provided the latter is performed over an energy interval
containing a large number of CN resonances with a smooth averaging
function $|f(E)|^2$ as in Eq.~(\ref{2}). The average $S$--matrix
$\langle S_{b a} \rangle$ describes the fast part of the reaction.
Simple models like the optical model for elastic scattering or
direct--reaction models yield reliable theoretical results for
$\langle S_{b a} \rangle$. The fluctuating part $S^{\rm fl}_{b a}(E)$
describes the slow part of the reaction, i.e., formation and
subsequent decay of the CN resonances.  Because of the complexity of
the latter, the precise energy dependence of $S^{\rm fl}_{b a}(E)$
cannot be predicted theoretically. Using a random--matrix model for
the CN resonances one can, however, calculate the
energy--autocorrelation function of $S^{\rm fl}_{b a}(E)$. That
function is given in terms of the average spacing $d$ of the CN
resonances and of the elements of the average $S$--matrix $\langle S
\rangle$ which serve as input parameters. We first assume that
$\langle S \rangle$ is diagonal and later discuss modifications due to
the non--zero non--diagonal elements. We have~\cite{Ver85}
\begin{align}
\label{5}
&\langle S^{\rm fl}_{b a}(E_1)(S^{\rm fl}_{b a}(E_2))^* \rangle =
\langle S^{\rm fl}_{b a}(E - \ve/2)(S^{\rm fl}_{b a}(E + \ve/2))^*
\rangle \nonumber \\
&=\prod_{i = 1}^2 \int_{0}^{+ \infty} {\rm d} \lambda_i \int_{0}^{1}
{\rm d} \lambda \ J_{b a}(\lambda_1, \lambda_2, \lambda)
\nonumber \\
&\times\frac{1}{8}\mu(\lambda_1,\lambda_2,\lambda)\
\exp\bigg\{ -\frac{i \pi\ve}{d}(\lambda_1+\lambda_2+2\lambda)\bigg\} 
\nonumber \\ 
&\times\prod_e\frac{(1 - T_e
\lambda)}{(1 + T_e \lambda_1)^{1/2} (1 + T_e \lambda_2)^{1/2}} \, .
\end{align}
We note that the autocorrelation function~(\ref{5}) depends only on
the energy difference $\ve = E_2 - E_1$. The factor $\mu(\lambda_1,
\lambda_2, \lambda)$ is an integration measure and is given by
\be
\mu(\lambda_1, \lambda_2, \lambda) = \frac{(1 - \lambda) \lambda
|\lambda_1 - \lambda_2|}{\prod_{i = 1}^2 [((1 + \lambda_i)
\lambda_i)^{1/2} (\lambda + \lambda_i)^2]} \, ,  
\label{6}
\ee
while
\begin{align}
\label{7}
&J_{b a}(\lambda_1, \lambda_2, \lambda) = (1 + \delta_{a b}) T_a T_b 
\\
&\times \bigg( \sum_{i = 1}^2 \frac{\lambda_i(1 + \lambda_i)}{(1 +
T_a \lambda_i)(1 + T_b \lambda_i)} + \frac{2 \lambda (1 - \lambda)}
{(1 - T_a \lambda) (1 - T_b \lambda)} \bigg) \nonumber \\
&\qquad + \ \delta_{a b} T^2_a (1 - T_a) \bigg( \sum_{i = 1}^2
\frac{\lambda_i}{1+ T_a \lambda_i} + \frac{2 \lambda}{1 - T_a
\lambda} \bigg)^2
\nonumber
\end{align}
describes the dependence of the correlation function on entrance and
exit channels $a$ and $b$. The product in Eq.~(\ref{5}) extends
over all open channels $e$. The correlation function~(\ref{5}) depends
on the average level spacing $d$ of the CN resonances and on the
``transmission coefficients''
\be
T_a = 1 - |\langle S_{a a} \rangle|^2 \, . 
\label{9}
\ee
Given $\langle S_{a a} \rangle$ and $d$, the right--hand side of
Eq.~(\ref{5}) is completely known. The full $S$--matrix
autocorrelation function reads then as
\be
C_{b a}(\ve) = \delta_{a b} |\langle S_{a a} \rangle|^2 + \langle
S^{\rm fl}_{b a}(E_1) (S^{\rm fl}_{b a}(E_2))^* \rangle \, .
\label{8}
\ee

We turn to the case where $\langle S \rangle$ is not diagonal. This is
of practical interest because the dipole operator gives rise to direct
$(\gamma_i, n)$ reactions and, thus, to non--vanishing elements
$\langle S_{n \gamma_i} \rangle$. We assume that $\langle S_{n
\gamma_i} \rangle$ and $\langle S_{n n} \rangle$ are known and that
$\langle S_{\gamma_i \gamma_j} \rangle$ is diagonal.  (Non--diagonal
contributions $\langle S_{\gamma_i \gamma_j} \rangle$ with $i \neq j$
would be of second order in the electromagnetic interaction and, thus,
negligible). We follow the work of Ref.~\cite{Eng74} summarized in
Ref.~\cite{Mit10}. If $\langle S_{a b} \rangle$ is not diagonal, one
has to determine the unitary transformation $U_{a b}$ that
diagonalizes the transmission matrix $P_{a b} = \delta_{a b} - \sum_c
\langle S_{a c} \rangle \langle S^*_{b c} \rangle$ so that $(U P
U^\dag)_{a b} = \delta_{a b} p_a$. The $S$--matrix transforms
according to $S \to \tilde{S} = USU^T$ where $\langle \tilde{S}
\rangle$ is also diagonal. The correlation function of $\tilde{S}$ is
given by Eqs.~(\ref{5}) to (\ref{7}), with all $T_a$ replaced by
$p_a$. Since $\langle S_{n \gamma_i} \rangle$ is governed by the
electromagnetic interaction, we have that $|\langle S_{n \gamma_i}
\rangle | \ll |\langle S_{n n} \rangle|$ for all $i$. This allows us
to use first--order perturbation theory to calculate $U$ and the
eigenvalues $p_a$. We find that to lowest non--vanishing order in the
electromagnetic interaction, the autocorrelation function of $S =
U^\dag \tilde{S} U^*$ is equal to that of $\tilde{S}$ as given in
Eqs.~(\ref{5}) to (\ref{7}), except for the replacement
\be
T_{\gamma_i} \to p_{\gamma_i} = T_{\gamma_i} - T_n |\langle S_{n
\gamma_i} \rangle|^2 \, .
\label{12}
\ee
The autocorrelation function is then
\be
C_{b \gamma_0}(\ve) = |\langle S_{b \gamma_0} \rangle|^2 + \langle
S^{\rm fl}_{b \gamma_0}(E_1) (S^{\rm fl}_{b \gamma_0}(E_2))^* \rangle
\, .
\label{13}
\ee
The term $\langle S^{\rm fl}_{b \gamma_0}(E_1) (S^{\rm fl}_{b
\gamma_0}(E_2))^* \rangle$ is given by Eqs.~(\ref{5}) to (\ref{7})
with the replacement Eq.~(\ref{12}).

We apply these results to the case of interest by choosing $a =
\gamma_0$ and calculating the Fourier transform of $C_{b
\gamma_0}(\ve)$ and, from there, the intensity $I^{\rm long}_{b
\gamma_0}(t)$. We find (see Eq.~(\ref{3}))
\begin{align}
\label{10}
& \frac{k^2_0}{\pi} I^{\rm long}_{b \gamma_0}(t) = 2 \pi \hbar
\delta(t) \delta_{b \gamma_0} |\langle S_{\gamma_0 \gamma_0}
\rangle|^2 \\
& + \int {\rm d} \ve \ \exp \{ i \ve t / \hbar \} \langle S^{\rm
fl}_{b \gamma_0}(E - \ve/2) (S^{\rm fl}_{b \gamma_0}(E +
\ve/2))^*\rangle\, . \nonumber
\end{align}
The delta function in the first term on the right--hand side of
Eq.~(\ref{10}) signals that the contribution from $\langle S \rangle$
is instantanteous in time. That would hold for an infinitely short
laser pulse. For the actual laser pulse, the signal will have the same
time duration $\Delta t$ as the pulse itself. In any case, that term
does not contribute to the long--term behavior. Thus, 
\begin{align}
\label{10a}
& \frac{k^2_0}{\pi} I^{\rm long}_{b \gamma_0}(t) = {\cal C}_{b
\gamma_0}(t) \\
& = \int {\rm d} \ve \ \exp \{ i \ve t / \hbar \} \langle S^{\rm
fl}_{b \gamma_0}(E - \ve/2) (S^{\rm fl}_{b \gamma_0}(E + \ve/2))^*
\rangle \, . \nonumber
\end{align}
The first equation defines ${\cal C}_{b \gamma_0}(t)$. From
Eq.~(\ref{5}) we have
\begin{align}
\label{11}
& {\cal C}_{b \gamma_0}(t) 
\\
&= 2 d \prod_{i = 1}^2 \int_{0}^{+ \infty} {\rm d}
\lambda_i \int_{0}^{1} {\rm d} \lambda \ \delta( [ t d / (\pi \hbar) ] -
[\lambda_1 + \lambda_2 + 2 \lambda ] ) 
\nonumber \\
&\times \frac{1}{8} \ \mu(\lambda_1, \lambda_2, \lambda)
\prod_{i = 0}^\Lambda \frac{(1 - T_{\gamma_i} \lambda)}{(1 +
T_{\gamma_i} \lambda_1)^{1/2} (1 + T_{\gamma_i}
\lambda_2)^{1/2}} 
\nonumber \\
&\times ( [ 1 - \delta_{n, {\rm open}}] +  \delta_{n, {\rm
open}} \frac{(1 - T_n \lambda)}{(1 + T_n \lambda_1)^{1/2}
(1 + T_n \lambda_2)^{1/2}} ) 
\nonumber \\
&\times J_{b \gamma_0} (\lambda_1, \lambda_2, \lambda) \, .
\nonumber 
\end{align}
where the replacement Eq.~(\ref{12}) has to be made. The function
${\cal C}_{b \gamma_0}(t)$ gives the decay intensity of the CN resonances and
is the object of central interest. The delta function under the
integral in Eq.~(\ref{11}) shows that contributions due to the decay
of the CN resonances are delayed (all three integration variables are
positive). The symbol $\delta_{n, {\rm open}}$ is zero (unity) if the
neutron channel is closed (open), respectively.

\section{Decay in Time of the CN Resonances}

In this and the next Section we work out the time dependence of the
time--decay function ${\cal C}_{b \gamma_0}$ defined in Eqs.~(\ref{10a}) and
(\ref{11}), with the replacement~(\ref{12}), in a time domain where
the signal is sufficiently strong for detection. The average
correlation width $\Gamma$ of the CN resonances is approximately given
by the Weisskopf estimate~\cite{Bla52}, $\Gamma = (d / (2 \pi)) \sum_e
p_e$. Since the CN resonances are isolated, we have $\sum_{i =
0}^\Lambda p_{\gamma_i} < 1$. The number $(\Lambda + 1)$ of open gamma
channels is very large, $\Lambda \gg 1$, so that $p_{\gamma_i} \ll 1$
individually for all $i$. The value of the transmission coefficient
$p_{\gamma_0}$ can be obtained from the average total cross section
for dipole absorption given by
\begin{equation}
\sum_b \langle \sigma_{b \gamma_0} \rangle = \frac{2 \pi}{k^2_0} [ 1 -
\Re \langle S_{\gamma_0 \gamma_0} \rangle ] \, .
\label{19}
\end{equation}
According to the statistical model, $\langle S_{\gamma_0 \gamma_0}
\rangle$ is real and, for weak coupling to the channels, positive.
From Eq.~(\ref{9}) (with $T$ replaced by $p$) we have $\langle
S_{\gamma_0 \gamma_0} \rangle = \sqrt{1 - p_{\gamma_0}}$ and, for
$p_{\gamma_0} \ll 1$, $\langle S_{\gamma_0 \gamma_0} \rangle \approx 1
- p_{\gamma_0} / 2$. That yields
\begin{equation}
p_{\gamma_0} \approx \frac{k^2_0}{\pi} \sum_b \langle \sigma_{b
\gamma_0} \rangle \, .
\label{20}
\end{equation}

For the transmission coefficient $T_n$, we observe that the neutron
has angular momentum one (zero) if the parity of the residual nucleus
is positive (negative), respectively. We use the fact that for the CN
resonances seen in slow $s$--wave neutron scattering, the average
width $\Gamma$ is about $1$ eV. Moreover, $\Gamma$ is dominated by the
neutron channel, $\Gamma \approx \Gamma_n$. With $T_n = 2 \pi \Gamma_n
/ d$ and $d \approx 10$~eV that gives $T_n \approx 0.6$ for $s$--wave
neutrons.  The transmission coefficient for $p$--wave neutrons is
smaller by the $p$--wave angular momentum barrier penetration factor
$kR$ (with $R$ the nuclear radius and $k$ the wave number). For a
laser pulse of several $10$~keV band width and a mean energy of $50$
keV above neutron threshold, we have $kR \approx 0.25$, leading to
$T_n \approx 0.1$ or $0.2$. At an energy right above neutron threshold
$T_n$ is considerably smaller, of course. Thus, $T_n$ is much larger
than any of the $p_{\gamma_i}$ and of the same order as or even larger
than $\sum_i p_{\gamma_i}$. This shows that we must treat $T_n$ and
$p_{\gamma_i}$ in Eq.~(\ref{11}) differently.

The evaluation of Eq.~(\ref{11}) seems to require the knowledge of all
individual transmission coefficients $p_{\gamma_i}$ for all photon
channels. These are not known. However, Eq.~(\ref{11}) can be much
simplified so that it depends only on the total decay width for gamma
decay and on the transmission coefficients in the entrance, in the
exit, and in the neutron channel.

We are guided by the following observation. In Ref.~\cite{Har92} it
was shown (see Eq.~(6.12) of that reference) that if $\sum_i T_i \ll
1$, the Fourier transform of the $S$--matrix autocorrelation function
$\langle S_{a b}(E - \ve / 2) S^*_{a b}(E + \ve / 2) \rangle$ is given
by
\be
{\cal C}_{b a}(t)\simeq
\frac{1}{\left[1 + T_a \frac{dt}{\pi\hbar}\right]
\left[1 + T_b \frac{dt}{\pi \hbar}\right]} 
\prod_c\frac{1}{\sqrt{1 + T_c\frac{dt}{\pi \hbar}}}\, . 
\nonumber
\ee
If the number of channels is large, that expression is approximately
given by $\exp [ - (d t / h) ( \sum_i T_i + 2 T_a + 2 T_b) ]$, and the
approximation is excellent for times $t \ll h / (d T_i)$ for all
$i$. If all $T_i$ are approximately equal, that condition is met for
all times for which the signal is detectable. That shows that for
$\Lambda \gg 1$, the non-exponential decay predicted in
Ref.~\cite{Har92} actually becomes unobservable: Deviations from the
exponential decay form occur only for times for which the signal is
too small for detection. We use that fact to simplify
Eq.~(\ref{11}). We assume that the $p_{\gamma_i}$ all have similar
values and use the approximation $(1 - p_{\gamma_i} \lambda) [ (1 +
p_{\gamma_i} \lambda_1) (1 + p_{\gamma_i} \lambda_2) ]^{- 1/2}
\approx \exp ( - [p_{\gamma_i}/2] [ \lambda_1 + \lambda_2 + 2 \lambda ]
)$. That implies
\be
\prod_{i = 0}^\Lambda \frac{(1 - p_{\gamma_i}
\lambda)}{(1 + p_{\gamma_i} \lambda_1)^{1/2} (1 + p_{\gamma_i}
\lambda_2)^{1/2}} \to \exp \{ - \sum_{i = 0}^\Lambda p_{\gamma_i} t d
/ h \} \, .
\label{14}
\ee
The arrow indicates that we have used the delta function in
Eq.~(\ref{11}). We reiterate that the approximation Eq.~(\ref{14}) is
expected to be excellent for times that allow a detection of the
signal although it may fail asymptotically ($t \to \infty$). To
interpret the right--hand side of expression Eq.~(\ref{14}) we use that
for $T_{\gamma_i} \ll 1$ we have $T_{\gamma_i} = 2 \pi
\Gamma_{\gamma_i} / d$ so that $(d / h) \sum_i T_{\gamma_i} =
\Gamma_\gamma / \hbar$ where $\Gamma_{\gamma_i}$ and $\Gamma_\gamma$
are the average partial and total widths for gamma decay of the CN
resonances, respectively. That is the expected result. The replacement
Eq.~(\ref{12}) implies that the average partial widths and the total
width for gamma decay are reduced by the direct reaction. This is
plausible because some decay strength is taken away by that reaction.
We denote the ensuing average total width for gamma decay by
$\tilde{\Gamma}_{\gamma}$. The corresponding increase in neutron decay
width is negligibly small in comparison with $T_n$.

With the approximation leading to expression~(\ref{14}),
Eq.~(\ref{11}) becomes
\begin{align}
\label{15}
&\int {\rm d} \ve \ \exp \{ i \ve t / \hbar \} \langle S^{\rm fl}_{b
\gamma_0}(E - \ve/2) (S^{\rm fl}_{b \gamma_0}(E + \ve/2))^* \rangle 
\nonumber \\
&= 2 d \exp \{-\tilde{\Gamma}_\gamma t / \hbar \} F_{b \gamma_0}(t)
\end{align}
where
\begin{align}
\label{16}
&F_{b \gamma_0}(t) 
\\
&=\prod_{i = 1}^2 \int_{0}^{+ \infty} {\rm d}
\lambda_i \int_{0}^{1} {\rm d} \lambda \ \delta( [ t d / (\pi \hbar)
] - [\lambda_1 + \lambda_2 + 2 \lambda ] ) 
\nonumber \\ 
&\times \frac{1}{8} \ \mu(\lambda_1, \lambda_2, \lambda) 
\nonumber \\
&\times ( [ 1 - \delta_{n, {\rm open}}] + \delta_{n, {\rm
open}} \frac{(1 - T_n \lambda)}{(1 + T_n \lambda_1)^{1/2} (1 + T_n
\lambda_2)^{1/2}} ) 
\nonumber \\
&\times J_{b \gamma_0} (\lambda_1, \lambda_2, \lambda) \, .
\nonumber
\end{align}
The time--decay function in Eq.~(\ref{15}) is the product of two
factors. The first is an exponential and describes the decay due to
all gamma transitions (except for additional contributions from the
entrance and the exit channels). The second factor $F_{b
\gamma_0}(t)$, given in Eq.~(\ref{16}), depends on the channels under
consideration and on a small number of parameters: Via Eq.~(\ref{7})
$F_{b \gamma_0}(t)$ depends on the transmission coefficients in the
entrance and exit channels and via the delta function on the average
level spacing of the CN resonances. Moreover, $F_{b \gamma_0}(t)$
depends on whether the neutron channel is open or not, and on the
value of $T_n$.  Interest focuses on $F_{b \gamma_0}(t)$ because it
gives rise to observable modifications of the exponential decay.

We expect $F_{b \gamma_0}(t)$ to vanish for $t \leq 0$ (this is
confirmed by the delta function in Eq.~(\ref{16})), to rise to a
maximum at some positive value of $t$, and to decay towards zero for
$t \to \infty$.  Obvious questions are: At which value of $t$ does the
maximum of $F_{b \gamma_0}(t)$ occur? How steep is the rise for small
positive values of $t$?  What is the form of the decay for values of
$t$ beyond the maximum? Some of these questions can be answered
analytically. For small positive times, $F_{b \gamma_0}(t)$ rises
quadratically. Indeed, because of the delta function in the integrand
of Eq.~(\ref{16}) both $F_{b \gamma_0}(t)$ and its first derivative
vanish at $t = 0$. For larger values of $t$ the behavior of $F_{b
\gamma_0}(t)$ is expected to differ for the three possible cases: (i)
the neutron channel is closed, (ii) the neutron channel is open and
neutron decay is measured ($b = n$) and (iii) the neutron channel is
open but gamma decay is measured ($b = \gamma_i$). The results of
Ref.~\cite{Har92} suggest that in case (i) the peak of $F_{\gamma_i
\gamma_0}(t)$ is followed by a decay of the form $t^{- 2}$. However,
since both $p_{\gamma_0}$ and $p_{\gamma_i}$ are very small in
comparison with $\tilde{\Gamma}_\gamma$, that decay is very slow, and
the behavior of the time--decay function~(\ref{15}) beyond the peak of
$F_{\gamma_i \gamma_0}(t)$ is governed by the first factor in
Eq.~(\ref{15}), i.e., is purely exponential. In case (ii) we expect
that $T_n d / \hbar$ is at least as large as $\tilde{\Gamma}_\gamma$.
The large flux into the neutron channel should shift the peak of $F_{n
\gamma_0}(t)$ towards smaller values of $t$ than in case (i). The
decay in time of $F_{n \gamma_0}(t)$ beyond its peak should be
governed by the neutron channel, too, and should asymptotically be
proportional to $t^{- 3/2}$. The decay time is comparable with or
smaller than $\hbar / \tilde{\Gamma}_\gamma$, and modifications of the
exponential decay form should be detectable. Similar statements apply
in case (iii) except that now beyond its peak the time decay of
$F_{\gamma_i \gamma_0}(t)$ is asymptotically given by $t^{- 1/2}$.

\section{Numerical Results}

\begin{figure}[ht]
        \centering \includegraphics[width=8.5cm]{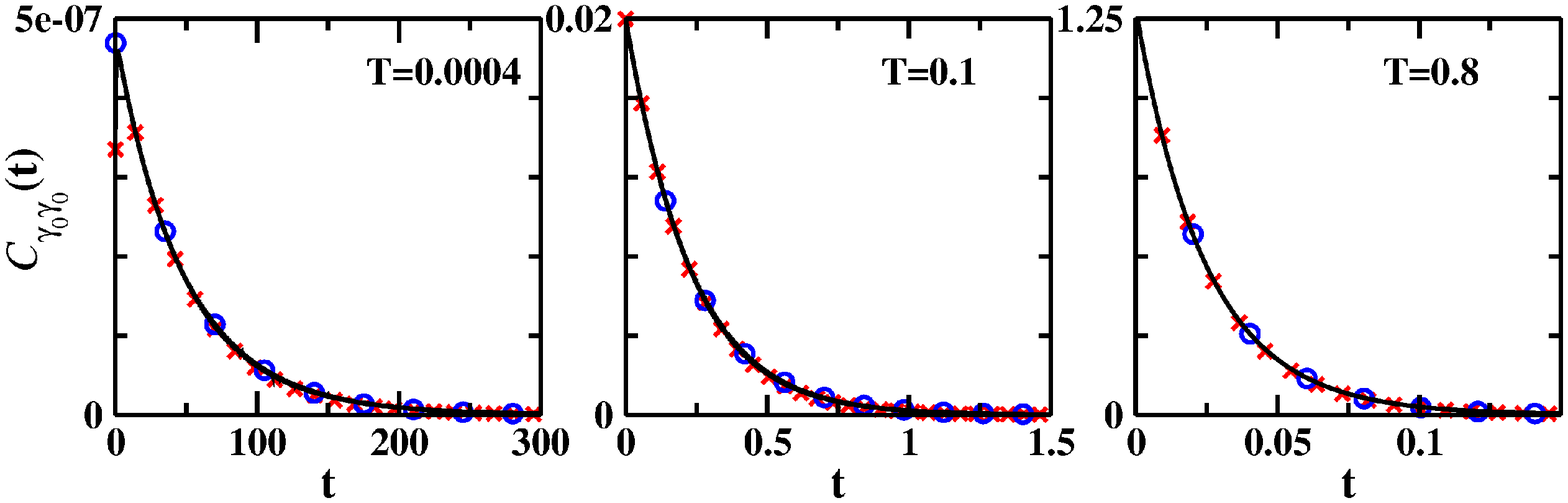}
        \caption{Intensity ${\cal C}_{\gamma_0 \gamma_0}(t)$ versus
        time $t$ in units of $h / d$. The neutron channel is closed,
        and $\Lambda = 49$ inelastic gamma channels are open, with
        $p_{\gamma_i} = p_{\gamma_0} = T$, $i = 1,...,49$ and values
        of the transmission coefficients chosen as indicated in the
        insets. Solid line (color online: black): Eq.~(\ref{11}).
        Crosses (color online: red): Approximation Eq.~(\ref{15}).
        Open circles (color online: blue): Fit of an exponential
        $\exp(-a_2t)$ to the data, resulting in $a_2 \approx \sum_{i
        = 0}^\Lambda p_{\gamma_{i}}$.}  \label{fig1}
\end{figure}

\begin{figure}[ht]
        \centering
        \includegraphics[width=8.5cm]{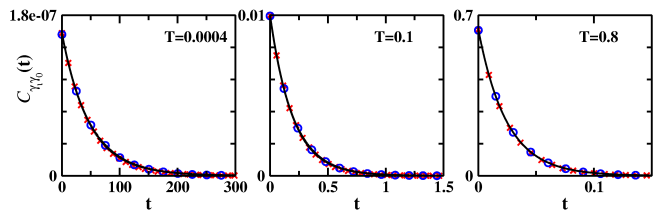}
        \caption{Same as Fig.~\ref{fig1} but for an inelastic gamma
        channel, ${\cal C}_{\gamma_i \gamma_0}(t)$ with $i \neq 0$.}
        \label{fig2}
\end{figure}

\begin{figure}[ht]
        \centering \includegraphics[width=8.5cm]{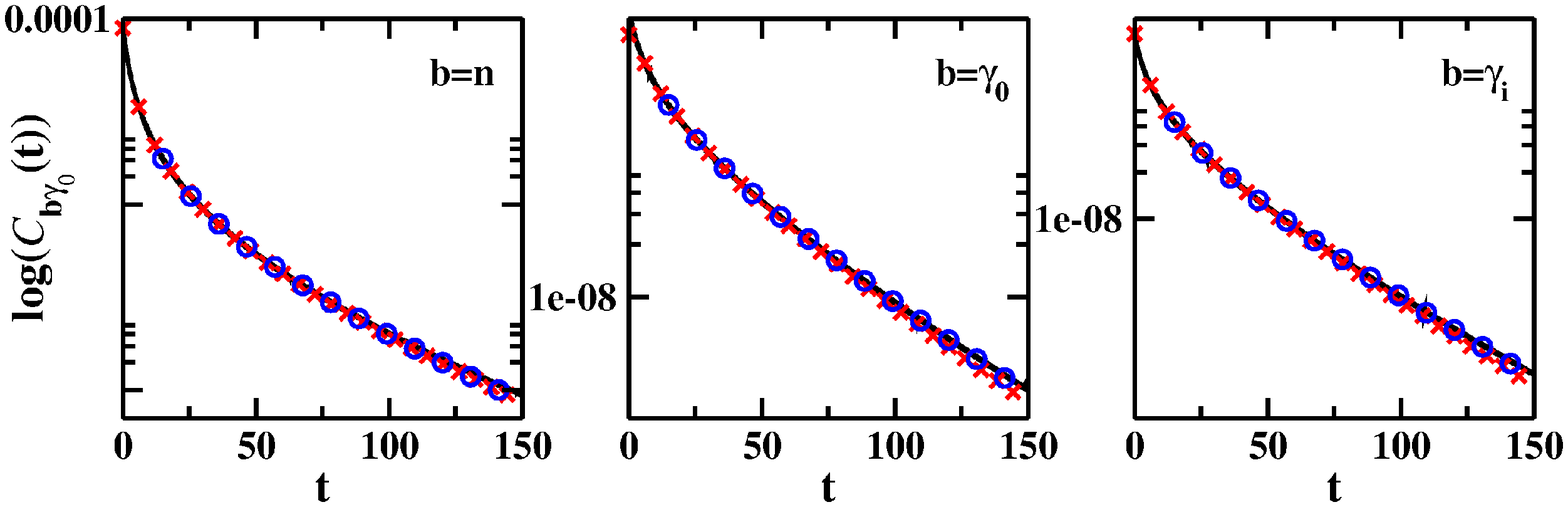}
        \caption{Logarithmic plot of ${\cal C}_{b \gamma_0}(t)$ versus
        $t$ (in units of $h / d$) for an open neutron channel and
        $\Lambda = 49$ inelastic gamma channels, with $p_{\gamma_i} =
        p_{\gamma_0} = T$, $i=1,...,49$, and values of the
        transmission coefficients $T=0.0004$, $T_n=0.2$ and of the
        final channel $b$ as indicated in the insets. Solid line
        (color online: black): Eq.~(\ref{11}) for large times $t$.
        Crosses (color online: red): Approximation Eq.~(\ref{15}).
        Open circles (color online: blue): Fit of $t^{a_1} \exp(
        -{a_2}t)$ to data. In all three cases $a_2 \simeq 0.02$ which
        is $\approx 50 \times T$. For $b = n$ we find $a_1 \simeq 1.46
        \approx 3/2$, for $b = \gamma_0, \ \gamma_i$ we have $a_1
        \simeq 0.48 \approx 1/2$.}
        \label{fig3}
\end{figure}

\begin{figure}[ht]
        \centering \includegraphics[width=8.5cm]{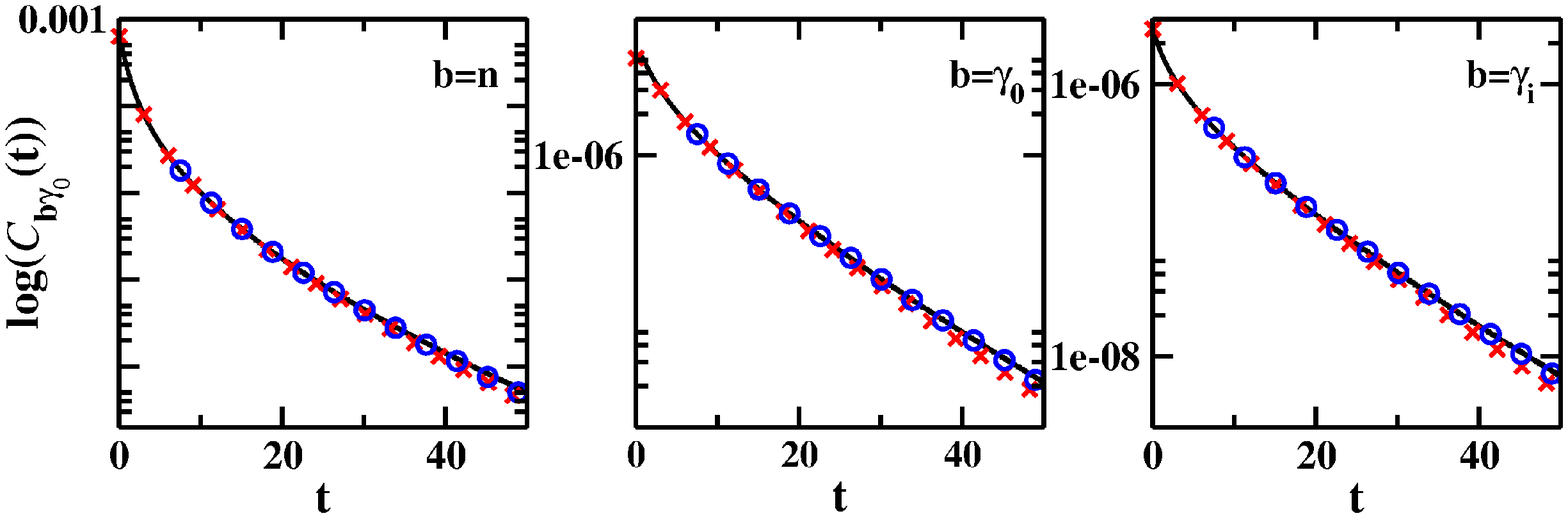}
        \caption{Same as Fig.~\ref{fig3} but for a different set of
        transmission coefficients $T_n = 0.4$, $T = 0.0016$. In all
        three cases $a_2 \simeq 0.077$ which is $\approx 50 \times
        T$. For $b = n$ we have $a_1 \simeq 1.39 \approx 3/2$, for $b
        = \gamma_0,\ \gamma_i$ we obtain $a_1 \simeq 0.51 \approx
        1/2$.}  \label{fig4}
\end{figure}

Further insight into the time dependence of the time--decay
function~(\ref{11}) is obtained by numerical simulation. Taking
$\Lambda = 49$, choosing all $p_{\gamma_i}$ equal to $T$, and using
different sets of values for $T$ and $T_n$, we have found that
Eq.~(\ref{15}) is in all cases an excellent approximation to
Eq.~(\ref{11}) for those values of $t$ for which the signal is
detectable, both when the neutron channel is closed and when it is
open. By way of example this is shown in Figs.~\ref{fig1} to
\ref{fig4}, with the black lines showing the analytic function
Eq.~(\ref{11}) and the red crosses the approximation
Eq.~(\ref{15}). That is an important result as it enables the analysis
of data in terms of a few parameters and without knowledge of the
individual values of the transmission coefficients in every gamma
channel.

The rise in time of the time--decay function ${\cal C}_{b
\gamma_0}(t)$ is very steep in all cases. The function reaches its
maximum at times of order $\hbar / d$. The maximum value of ${\cal
C}_{b \gamma_0}$ is mainly determined by the factor $T_a T_b$ in
Eq.~(\ref{7}); that explains the enormous difference in scale in the
figures. With increasing values of the transmission coefficients the
maximum is shifted towards smaller values of $t$. These are very short
times; it is not clear whether in the first round of experiments the
decay signal can be clearly separated from the signal due to the short
pulse itself (delta function in Eq.~(\ref{10})). Therefore, we have
focussed attention on the decay in time of ${\cal C}_{b \gamma_0}(t)$
beyond its maximum.

When the neutron channel is closed, ${\cal C}_{b \gamma_0}(t)$ is very
well approximated by an exponential, as expected. This is shown in
Figs.~\ref{fig1} and \ref{fig2} with the exponential fit shown as blue
open circles but applies equally to all other cases calculated.  The
exponential is the same for all gamma channels. This is important
because the signal is expected to be very weak for every single gamma
channel. Summation over many such channels does not affect the form of
the exponential. Measurements of a restricted sum $\sum_i {\cal
C}_{\gamma_i \gamma_0}(t)$ would yield the average gamma width of the
CN resonances.

Taken by itself, exponential decay is expected and not very exciting.
The situation changes when the neutron channel is open. This is shown
in Figs.~\ref{fig3} and \ref{fig4}. Because of the comparatively large
value of $T_n$ the neutron yield is significantly larger than the
yield in any single gamma channel. Moreover, the time decay function
differs significantly from an exponential, both in the neutron and in
the gamma channels. As indicated in the captions, the curves were
fitted with a function of the form $t^{a_1} \exp(-{a_2}t)$ (blue open
circles). The best--fit value of $a_2$ is approximately equal to the
sum of the transmission coefficients of the gamma channels. The
exponent $a_1$ agrees approximately with the result given at the end
of the last Section. Both results are in agreement with our
expectations.

\section{Summary and Conclusions}

In nuclear reactions induced by short laser pulses of several MeV
energy, the observable of interest is the time--decay function of the
CN resonances. Provided the length $\Delta t$ of the laser pulse is
chosen optimally, the time--decay function is given by the Fourier
transform of the $S$--matrix autocorrelation function. For that to be
true, $\Delta t$ must be large compared to the Heisenberg time $\hbar
/ d$ and small compared to the width of the giant dipole resonance.
This fixes $\Delta t$ to values between $0.5 \times 10^{- 19}$ s and
$0.5 \times 10^{- 20}$ s, depending on mass number. The time--decay
function comprises information on amplitude correlations of CN
resonances which cannot be obtained from other observables.

We have calculated the time--decay function under the assumption that
the laser--excited CN resonances are described by random--matrix
theory. Our Eq.~(\ref{15}) gives an excellent approximation to that
function. It depends on a small number of parameters only and is
useful for the analysis of data. We have shown how to estimate these
parameters from existing data (transmission coefficient for neutrons,
average cross section for dipole absorption). The time--decay function
rises steeply with time and reaches a maximum a short time after the
initial laser pulse has hit the target. That time is of the order of
the Heisenberg time. The further development in time of the
time--decay function depends on whether the neutron channel is closed
or open. In the first case, the time--decay function decreases
exponentially. The decay width is given by the average total gamma
decay width of the CN resonances and can be determined from data. In
the second case, the time--decay function can be fit by the product of
an exponential (again determined by the average total width for gamma
decay) and a power law. The exponent of the latter depends on whether
the final channel is the neutron channel or a gamma channel.

An experimental confirmation of our predictions would establish a new
unambiguous test of random--matrix theory in nuclei. In addition, it
would make it possible to measure the average total width for gamma
decay of CN resonances located below neutron threshold.

One of us (HAW) is grateful to P. Thirolf for drawing his attention to
the problem, and for discussions. We are grateful to H. L. Harney,
T. Papenbrock, and A. Richter for a reading of the manuscript and
helpful comments. This work was partly supported through SFB634 by the
DFG.

\end{document}